# Real-Time Prediction of BITCOIN Price using Machine Learning Techniques and Public Sentiment Analysis

S M Raju[a,*], Ali Mohammad Tarif[b]

[a,b]Computer Science, International Islamic University Malaysia, Gombak, Malaysia

## Abstract

Bitcoin is the first digital decentralized cryptocurrency that has shown a significant increase in market capitalization in recent years. The objective of this paper is to determine the predictable price direction of Bitcoin in USD by machine learning techniques and sentiment analysis. Twitter and Reddit have attracted a great deal of attention from researchers to study public sentiment. We have applied sentiment analysis and supervised machine learning principles to the extracted tweets from Twitter and Reddit posts, and we analyze the correlation between bitcoin price movements and sentiments in tweets. We explored several algorithms of machine learning using supervised learning to develop a prediction model and provide informative analysis of future market prices. Due to the difficulty of evaluating the exact nature of a Time Series(ARIMA) model, it is often very difficult to produce appropriate forecasts. Then we continue to implement Recurrent Neural Networks (RNN) with long short-term memory cells (LSTM). Thus, we analyzed the time series model prediction of bitcoin prices with greater efficiency using long short-term memory (LSTM) techniques and compared the predictability of bitcoin price and sentiment analysis of bitcoin tweets to the standard method (ARIMA). The RMSE (Root-mean-square error) of LSTM are 198.448 (single feature) and 197.515 (multi-feature) whereas the ARIMA model RMSE is 209.263 which shows that LSTM with multi feature shows the more accurate result.

## 1. Introduction

Bitcoin price prediction has been an active area of research for a long time. Bitcoin, as a pioneer within the blockchain monetary renaissance [1], [2] plays an overwhelming part in an entirety cryptocurrency market capitalization environment. Hence, it is the incredible interest of machine learning and data mining community to be able to: (I) predict Bitcoin price changes (II) grant experiences to get it what drives the Bitcoin instability and way better assess related dangers in cryptocurrency domain. Many researchers worked on machine learning algorithms and sentiment analysis from social media to find out the bitcoin stock market price prediction.

*Corresponding author, [a]*Email address:* s.raju@live.iium.edu.my
[b]*Email address:* ali.tarif@live.iium.edu.my

In our project, we used a time series model and deep learning to leverage machine learning technology to predict the real-time price of Bitcoin. However, machine learning literature is lacking verification of whether or not the stock evaluation strategies are legitimate for the cryptocurrencies, and if so, how they may be modified. That is what features want to be eliminated or introduced as a foundation for price prediction, whether current machine learning algorithms work for cryptocurrencies, and which technique yields the excellent outcomes.

In this paper, we check out these questions. Such evaluation is applicable given an extremely good quantity of attention that cryptocurrencies, particularly Bitcoin, are generating. Each people and big financial corporations are drawn to cryptocurrencies due to the transparency and anonymity that they offer to their users, in addition to their resistance to fraud because of the dispersed nature of the ledger statistics. Moreover, purchasing cryptocurrencies are promising in terms of making income, and ought to be of interest to buyers. In addition, to familiarizing themselves with industry tendencies and political and financial information, they can utilize machine learning models to decide whether to buy or promote cryptocurrency.

*1.1. Bitcoin*

Bitcoin is a global cryptocurrency and online payment system that is highly stable and secure. It is Peer-to-peer value transfer and transaction protocol. Bitcoin transactions are verified by network nodes, published on a public ledger. Highly stable and secure Decentralized verification (Blockchains). "Bitcoin" is the unit of account on this ledger.

Smallest unit: a satoshi representing 0.00000001 bitcoin, one hundred millionths of a bitcoin. In 2017, the Bitcoin price rose from $900 at the beginning of the year to nearly $20000 at the end of the year.

*1.2. Sentiment Analysis*

Sentiment analysis (SA) is the system of extracting the polarity of individuals' subjective opinions from plain normal language texts. Sentiment analysis includes characterizing opinions in text into categories like "positive" or "negative" or "neutral.

For instance, an evaluation on a website is probably extensively positive about a virtual digicam, however, be especially negative about how heavy it's far. Being able to become aware of this kind of facts in a scientific way offers the vendor a much clearer image of public opinion than surveys or recognition corporations do, because the information is created with the aid of the customer.

*1.3. Price Prediction*

The Bitcoin's price varies similarly to a stock albeit in another way. There are some algorithms used on stock market data for price prediction but the parameters affecting Bitcoin are distinctive. Therefore, it is essential to expect the price of Bitcoin in order that correct investment decisions can be made. The price of Bitcoin does now not rely on at the business events or intervening government in contrast to the stock market. Hence, to expect the value we feel it is essential to leverage machine learning technology to expect the rate of Bitcoin.

*1.4. Problem Statements*

1. Bitcoin is the most complex cryptocurrency which value change in every second.
2. Investing money for Bitcoin is more risk and less profit.

*1.5. Project Objectives*

1. To predict bitcoin price with maximum efficiency using LSTM and ARIMA.
2. To compare between ARIMA and LSTM to find which is the best efficient algorithm for predicting bitcoin price.
3. To ensure less risk and more profit for investors.

*1.6. Project Scope*

Today Bitcoin is a secure transaction system that has a valuable impact on capital. They are awarded under a restriction in which customers offer their computer authority to register and listing trades with the bitcoins. The purchase and sale of Bitcoins in different currencies is carried out in an alternative workplace where "purchase" or "sell" requests are placed in the ordered e-book. "Buy" or "bid" offers to talk about the purpose of purchasing certain Bitcoins measures at a few costs while "provide" or "ask" offers to talk about the expectation of providing certain Bitcoins measures at a certain cost. The change is ordered through the coordination of pricing requests from the arrangement of e-books to a valid exchange between customers and suppliers.

## 2. Literature Review

Bitcoin is currently a new technology and the world most expensive cryptocurrency thus there are some price prediction models available. Amjad et al. utilized the historical time series price data for price forecast and exchanging methodology [3] and Garcia et al. also appeared that the increments in opinion polarization and trade volume precede rising of Bitcoin prices [4].

Chen and Lazer [5] determined investment methodologies by observing and classifying the twitter feeds. Go et al. train the classifiers utilizing the dataset clarified by distant supervision and approve the classification performance [6]. Go et al. refer to the powerful paper by Pang et al. [7] where those analysts have set a standard for machine-learning based opinion analysis. Their approach is credited as one of the primary attempts at applying machine learning strategies to the issue of opinion analysis [7].

Some recent works focused on high-frequency trading and applying deep-learning techniques such as RNN for the prediction on time series data that have been tested dense, feed-forward networks as function model [8]. McNally [9] predicts the Bitcoin pricing process using machine learning techniques, such as recurrent neural networks (RNNs) and long short-term memory (LSTM) and compare results with those obtained using autoregressive integrated moving average (ARIMA) models.

From [10] a comparison between multi-layer perceptron MLP and non-linear autoregressive exogenous (NARX) model is made. They conclude that MLP can also be used for stock market prediction even though it does not outperform the NARX model in price prediction. The authors made use of MATLAB's neural network toolbox to build and evaluate the performance of the network.

Another paper [11] deals with daily time series data 10-minute and 10-second time-interval data. They have created three time-series data sets for 30 60 and 120 minutes followed by performing GLM/random forest on the datasets which produce three linear models. These three models are linearly combined to predict the price of bitcoin.

According to [12] the author is analyzing what has been done to predict the U.S. stock market. the conclusion of his work is the mean square error of the prediction network was as large as the standard deviation of the excess return. However, the author is providing evidence that several basic financial and economic factors have predictive power for the market excess return.

In [13] instead of directly forecasting the future price of the stock the authors predict the trend of the stock. The trend can be considered as a pattern. They perform both short-term predictions day or week predictions and also long-term predictions months they found that the latter produced better results with 79% accuracy. Another interesting approach the paper reflects is the performance evaluation criteria of the network. Based on the predicted output the performance evaluation algorithm decides to either buy and sell or hold the stock.

In this paper, we explored some of the relevant methods of bitcoin sentiment prediction using tweets and Reddit posts and our approach is parametric and stems from a hypothetical modeling system based on stationarity and mixing.

## 3. Data Collection & Preprocessing

In this section, we have two phases relevant to data collection and preprocessing. Two different datasets were collected during the study; the first consisting of real-time Bitcoin price data and the other of tweets from Twitter and Reddit posts. These datasets were collected using API services from the dedicated server and allowing uninterrupted continuously real-time data.

### 3.1. Bitcoin Data in Real-Time

For the first phase of this paper, we have collected Bitcoin values from four different databases: Coinmarketcap, Bitstamp, Coinbase, and Blockchain Info. We gathered Bitcoin historical and real-time price data using their publicly available API. From Coinmarketcap and Blockchain Info, we have collected 11 key features (see Table 1) relevant to the Bitcoin values for our research with their real-time feature (e.g. created date). In addition, from Bitstamp database, we also acquired 10 more transaction-based different level of details which are returned by API call for the pricing data (see Table 2). This dataset was gathered an interval length of every one-minute and it continued the data collection process. From Coinbase API we collected bitcoin real-time data in order to predict the fluctuation of the bitcoin price with others collected dataset. We collected bitcoin price data by building an automated web scraper which is real-time data pulled from Coinbase API over the course of weeks or months or years depending on program continuation. All these key features are connected with bitcoin network and important to real-time observation of moving prices.

Table 1: Bitcoin key features

| Features | Databases | Descriptions |
|---|---|---|
| Price USD | Coinmarketcap | Bitcoin price in USD |
| 24h Volume USD | Coinmarketcap | Bitcoin trade volume in USD |
| Market Cap USD | Coinmarketcap | Market capital of bitcoin price |
| Available Supply | Coinmarketcap | Number of coins in existence available to the public |
| Total Supply | Coinmarketcap | Total number of coins in existence available to the public |
| Percentage Change 1h | Coinmarketcap | The percentage changes of bitcoin price in 1 hour |
| Percentage Change 24h | Coinmarketcap | The percentage changes of bitcoin price in 24 hours |
| Percentage Change 7d | Coinmarketcap | The percentage changes of bitcoin price in 7 days |
| USD Sell | Blockchain Info | Price of bitcoin sell on the day |
| USD Buy | Blockchain Info | Price of bitcoin buy on the day |
| USD 15m | Blockchain Info | Changes in bitcoin prices within 15 minutes |

Table 2: Bitstamp API features

| Features | Definitions |
|---|---|
| High | The highest price of bitcoin on that one-minute time |
| Last | The last price of bitcoin on that one-minute time |
| Timestamp | Timestamp server of bitcoin on that one-minute time |
| Bid | The highest price pays for a bitcoin on that one-minute time |
| VWAP | Volume-weighted average price trade on that one-minute time |
| Volume | Bitcoin volume on that one-minute time |
| Low | The lowest price of bitcoin on that one-minute time |
| Ask | The lowest price ask the seller for a bitcoin on that one-minute time |
| Open | The open price of bitcoin on that one-minute time |
| Datetime | Current date and time with on that one-minute time |

### 3.2. Tweets Data in Real-Time

Another part of the first phase is to collect data for the sentiment analysis using Twitter's streaming API was used in combination with Tweepy. Tweepy is an open source framework written in Python, facilitates tweet collection from Twitter's API. Tweepy allows for filtering based on hashtags or words, and as such was considered as an efficient way of collecting relevant data. The filter keywords were chosen by selecting the most definitive Bitcoin-context words, for example, "bitcoin" could include sentiments towards bitcoin, and so the scope must be tightened further to only include Bitcoin synonyms. These synonyms include: 'bitcoin', 'BTC'.

#### 3.2.1. Tweets Preprocessing

Tweets consist of many acronyms, emoticons and unnecessary records like images and URL's. So, tweets are preprocessed to symbolize accurate feelings of the public. For preprocessing of tweets, we employed 3 ranges of filtering: Tokenization, Stopwords elimination, and Regex matching for removing special characters.
1. Tokenization: Tweets are split into character words primarily based on the gap and irrelevant symbols like emoticons are removed. We shape a list of individual words for each tweet.
2. Stopword elimination: Words that don't explicit any emotion is called Stopwords. After splitting a tweet, words like a, is, the, with etc., are eliminated from the listing of phrases.
3. Regex matching for removing special characters: Regex matching in Python is completed to suit URLs and are changed through the time period URL. Regularly tweets consist of hashtags (#) and @ addressing other users. They may be additionally changed definitely. For instance, #Microsoft is replaced with Microsoft and @Billgates is changed with User. Extended word showing extreme feelings like cooooooooool! is changed with cool! After those tiers, the tweets are ready for sentiment classification.

#### 3.2.2. Tweets Sentiment Analysis

Tweets are classified based on the sentiment as Positive (polarity >0), Negative (polarity <0) and Neutral (polarity =0). For individually tweet sentiment score, we used Textblob to automatically be passed the tweet text for analysis sentiment and gives polarity score. Besides that, we also used Haven OnDemand, an API service for analysis sentiment from the tweets automatically. Both methods were used for Twitter and Reddit tweets sentiment analysis purposes.

### 3.3. The Merging of Datasets

In this phase, we combine two different datasets into single dataset by using merge data program which used Pandas library. Merged dataset had only price, sentiment, and time features in our case.

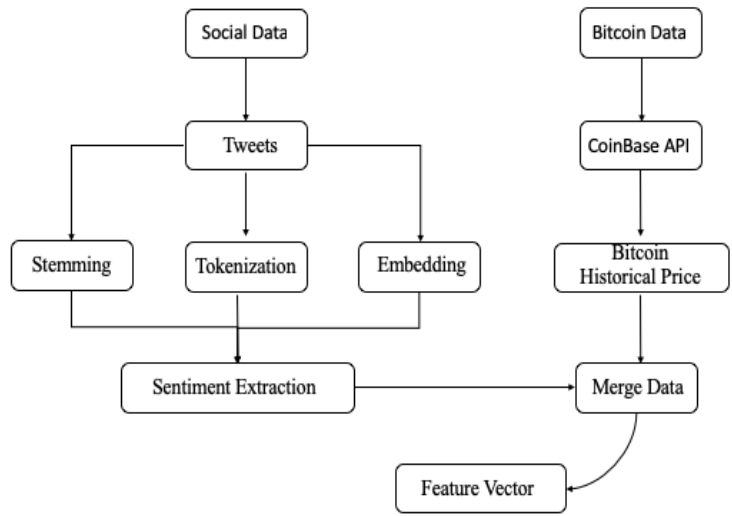

Fig. 1. Data Preprocessing

## 4. Methodology

*4.1. LSTM*

The long short-term memory network or LSTM addresses the common problem of disappearing gradients in the recurrent neural network. This is a type of recurrent neural network that is used in profound learning, as very large architectures can be trained. LSTM enables the network to learn more about many time steps by maintaining a more-steady error. This enables the network to learn long-term trust. LSTM cell contains forget and remember gates that allow the cell to decide which information to block or transmit based on its strength and importance [9]. As a result, weak signals that prevent the gradient from disappearing can be blocked. The performance of the RNN and LSTM network is assessed to determine the model's efficiency.

Elman's recent development of recurrent neural networks has gained popularity in network designs and increased computational power from graphical processing units. They are primarily useful with sequential data (in our case Bitcoins time series data) because each neuron or unit can access its internal memory to keep information about the previous input. Fig 2 shows simple RNN Structure [14]. One limitation of RNN is that it is influenced by the disappearing problem of the gradient. This problem is that since the layers and time steps of the network are interrelated, they are susceptible to exploding or disappearing gradients.

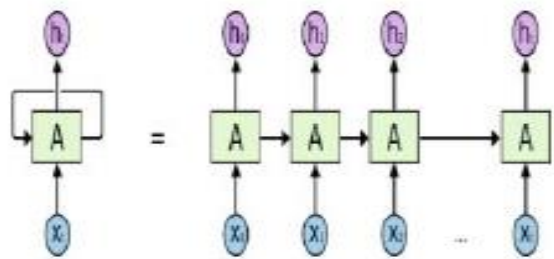

Fig. 2. RNN Structure

Vanishing gradients are a problem because they can prove to be too little to learn for the system, while inclinations can be restricted by regularization. In addition, some examinations have found that while RNN is able to deal with long-term dependencies, they often fail to learn in practice due to difficulties between gradient succession and long-term dependency.

*4.2. ARIMA*

The basic principle of the ARIMA model is to estimate the trend and the seasonality of the series and to remove them from the series in order to obtain a stationary series. In this series, statistical forecasting techniques can be used. The final step would be to convert the forecast values to the original scale by applying constraints on trend and seasonality.

Trend – mean varying over time. For example, in this case, we saw an average increase in the number of bitcoin price over time.

Seasonality – time frame variations. For example, people may tend to buy cars in a given month due to pay increments or festivals.

Static forecasts are performed and the RMSE is calculated to compare with LSTM model results. ARIMA model is implemented to compare its predictability with the LSTM and figure out which is the most suitable method for time series data which has huge fluctuations. ARIMA (Auto regression integrated moving average) is a class model that captures a suite of different standard temporal structures in time series data which include trend, seasonality, cycles, errors and nonstationary data. This allows it to exhibit dynamic temporal behavioral a time sequence. The data preparation phase is done similar to the LSTM model approach.

## 5. Implementation

The study focuses primarily on the Bitcoin closing price for the development of the predictive model. The increase or decrease in the Bitcoin price with the higher volatility makes it harder to predict, but the machine learning models try to predict with some degree of accuracy. The implementation here is performed using the LSTM Recurring Neural Network.

*5.1. Data Loading & Preparation*

The data is loaded from the .csv file using the panda library function read_csv(). The missing values are identified, and it is prepared for modeling by removing the unused fields and filling the NAN values.

*5.2. Data Normalization*

We need to normalize inputs in neural networks and other data mining models, otherwise, the network will perform poorly. Normalization is carried out in order to have the same range of values for each RNN model input. This can guarantee stable weight and partiality convergence. Normalization here uses the MinMaxScalar Package, after normalization, data is plotted using matplot libraries and the trend is seen to check the fluctuations in the price and volume of the bitcoin over the last 2 years (2017-2018).

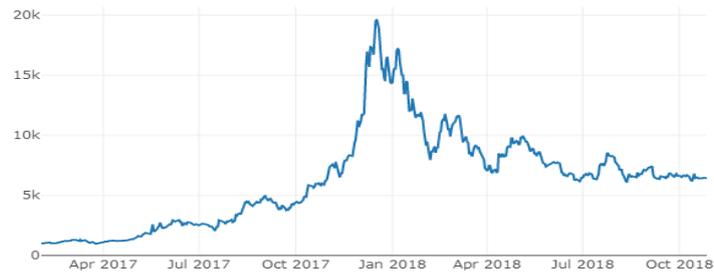

Fig. 3. Normalization

The above figure depicts that higher fluctuation in the bitcoin price from November 2017 to February 2018.

### 5.3. The Splitting of Data

We divide the data into training and test data in statistics and machine learning. The model is intended to fit the training data to make predictions. When we do this, there are chances of two things happening, one overfitting the model and the other underfitting the model. Overfitting means that the model is too well trained, and the predictions are too close and that it does not fit closely with the model when the model is underfitting. The Scikit Library is used here to split the data. The data for training is divided into 70 percent and 30 percent for testing. Total observation in the training dataset is 443 and 191 in the test dataset.

### 5.4. LSTM Modeling

The help of LSTMs preserves the error which can be reproduced over time and layers. By maintaining a more constant error, recurrent networks can continue to learn over a long period of time (over 200), opening a channel to remotely link causes and effects. S. Saravanakumar, V. Dinesh Kumar [15] LSTMs contain data outside the normal flow of the gated cell recurrent network. Information can be stored, written to or read from a cell in a computer memory similar to data. The cell decides what to store and when to read, write and erase through gates that open and close. The function below converts the series to supervised data. Two models are available in Keras. One is a sequential model that is suitable for predicting time series and the other is used with a functional API. The dense layer is used with input shapes as the output layer. The optimizer function used is 'adam' which has a learning rate of 0.01 with the mean absolute error as the loss function. The loss method used is mean absolute error.

### 5.5. The Root Mean Squared Error

Time series usually focus on predicting real values, which are called regression problems. Hence, the performance measurements in this tutorial focus on real-value prediction evaluation methods. The most commonly used mean squared error(MSE), root mean squared error(RMSE) etc.

*5.6. ARIMA Modeling*

ARIMA model uses and decomposes historical data into Autoregressive (AR) Indicates weighted average movement over past observations, Integrated (I) Indicates linear trends or polynomial trends and Moving Average (MA) Indicates weighted average movement over past mistakes. Therefore, the model has three parameters AR(p), I(d) and MA(q) all combined to form ARIMA (p, d, q) where

p= autocorrelation order
d= integration order (differentiation)
q= moving averages

Ref. [16] A non - seasonal stationary time series may be modeled as a combination of past values and errors known as ARIMA (p, d, q) or expressed as ARIMA (p, d, q).

To fit an ARIMA model that assumes stationary characteristics, we must use our data to determine the three parameters: p;d;q. p corresponds to the autoregressive component, q corresponds to the moving average and d corresponds to the degree of Box and Reinsel [17]; Robert H. Shumway [18].

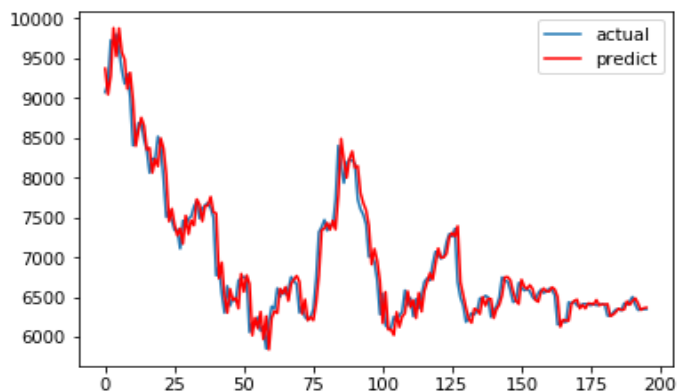

Fig. 4. ARIMA prediction result

The above figure shows the ARIMA point predictions (10,1,0). The model is updated with each new data point and the forecasts are displayed in the plot. The plots are zoomed in to show a few points to emphasize how the forecasts tend to be almost one-step versions of the original version. This does not seem to be very promising since all forecasts track the latest data point in the past and provide little new information. They also do not provide a likelihood for each forecast. This could be a serious handicap in combination with a trading strategy that could benefit from a probabilistic framework to determine the confidence with which it can engage in new trades.

## 6. Discussion and Results

In order to decide proper strategy of analysis for studying the connection amongst Bitcoin's price and others significant parameters as well as sentiment analysis, the available associated literature has been tested in depth. Maximum of articles [19], [20], [21] reports analysis about the existent relationship among the volume of tweets and the market evolution. In general, Bollen et al. demonstrated that tweets are expecting the market trend 3-4 days in advance, with a very good chance of success.

From our experiments, the result shows that machine learning models (LSTM) take much longer to compile because of their complex calculations than traditional models(ARIMA). The LSTM model compilation time is 61 milliseconds and 4 milliseconds for the ARIMA model [22].

The loss for the LSTM model is minimal at the learning rate of 0.01. It is not the best fit because it is almost impossible to meet both the train and the test at a time because the time series data fluctuates enormously. The lower loss makes the model better than the Bitcoin price forecast.

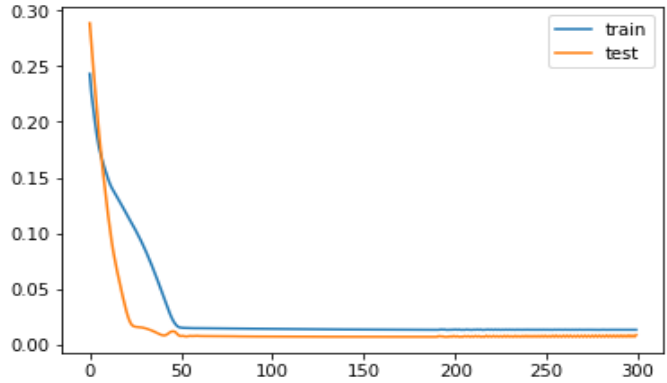

Fig. 5. Single train feature whereas the loss in y-axis and time points in the x-axis

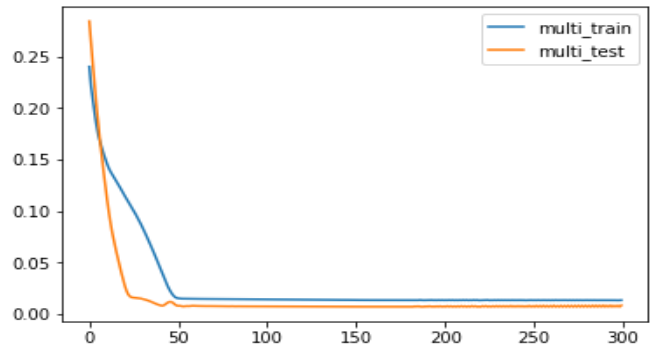

Fig. 6. Multi train feature whereas the loss in y-axis and time points in the x-axis

LSTM single feature and multi-feature predicting graphs are given below.

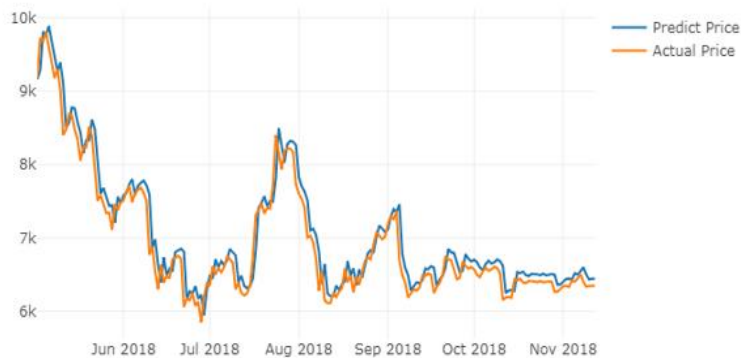

Fig. 7. LSTM single feature forecasting graph; price in y-axis and day in x-axis

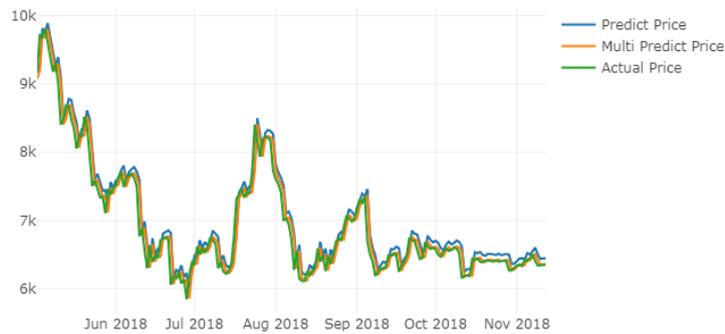

Fig. 8. LSTM multi-feature forecasting graph; price in y-axis and day in x-axis

The ARIMA model's RMSE is 209.263, while the LSTM RMSE is 198.448 (single feature) and 197.515 (multi- feature), which prove that the ARIMA model exceeds the machine learning algorithms in our LSTM model case. The Root Mean Square Error is minimal for the models because the data varies from 0 to 10000 USD and their closing price fluctuates enormously. Since the accuracy of the model in LSTM is minimal, the machine learning algorithm (Long Short Memory) is best suited for Bitcoin forecasting. LSTM is generally suitable for predicting higher fluctuations in the time series data.

## 7. Conclusion and Future Works

This study focuses on the Bitcoin closing price and sentiments of the current market for the development of the predictive model. It does also calculate the market sentiments to predict the price more accurately. The prediction is limited to previous data. The ability to predict data streaming would improve the model's performance and predictability. The model developed using LSTM is more accurate than the traditional models that demonstrate a deep learning model. In our case, LSTM (Long Short-Term Memory) is obviously an effective learner on training data than ARIMA, with the LSTM more capable of recognizing long-term dependencies. This study uses the daily price fluctuations of the Bitcoin to further investigate the model's predictability with hourly price fluctuations in the future. This paper consists only of comparing ARIMA with LSTM. The result would be confirmed by comparing more machine learning models in the future.

In this work, we have only considered Twitter and Reddit posts data to analyze people's feelings that may be biased because not all people who trade in stocks share their views on Twitter and Reddit. Moreover, Facebook posts and LinkedIn data can be included in a comprehensive collection of public opinion. In addition, the current sentiments can be combined with the prediction of the LSTM model to influence the decision of an autonomous trading assistant to buy or sell Bitcoins.

## 8. List of Abbreviations

USD — United States Dollar; RNN — Recurrent Neural Network; ARIMA — Autoregressive Integrated Moving Average; LSTM — Long Short-Term Memory, MSE — Mean-Square Error; RMSE — Root-Mean-Square Error; SA — Sentiment Analysis; DCI — Digital Currency Initiative; NARX — Non-linear Autoregressive Exogenous; MLP — Multi-layer Perceptron; GLM — Generalized Linear Model; API — Application Programming Interface; BTC — Bitcoin.

## Acknowledgments

Praise and thanks to Allah first and foremost whose blessing enabled us to accomplish this project. We wish to express our deepest appreciation to our supervisor Dr. RAINI HASSAN for continual guidance, helpful suggestion, close supervision and moral encouragement to complete this task.